\documentclass[preprint,floats,aps,epsfig,nofootinbib,amssymb,showpacs]{revtex4}
\usepackage{color}
\usepackage{graphicx}
\usepackage{tabularx}
\usepackage{txfonts}
\usepackage{amssymb}
\usepackage{dcolumn}
\usepackage{bm}

\begin{document}

\title{The real singlet scalar dark matter model}

\author{Wan-Lei Guo}
\email[Email: ]{guowl@itp.ac.cn}

\author{Yue-Liang Wu}
\email[Email: ]{ylwu@itp.ac.cn}

\affiliation{ Kavli Institute for Theoretical Physics China, \\
Key Laboratory of Frontiers in Theoretical Physics, \\
Institute of Theoretical Physics, Chinese Academy of Science,
Beijing 100190, China}

\begin{abstract}

We present an undated comprehensive analysis for the simplest dark
matter model in which a real singlet scalar with a $Z_2$ symmetry is
introduced to extend the standard model. According to the observed
dark matter abundance, we predict the dark matter direct and
indirect detection cross sections for the whole parameter space. The
Breit-Wigner resonance effect has been considered to calculate the
thermally averaged annihilation cross section. It is found that
three regions can be excluded by the current direct and indirect
dark matter search experiments. In addition, we also discuss the
implication of this model for the Higgs searches at colliders.

\end{abstract}


\maketitle

\section{Introduction}

The existence of dark matter (DM) is by now well confirmed
\cite{DM}. The recent cosmological observations have helped to
establish the concordance cosmological model where the present
Universe consists of about 73\% dark energy, 23\% dark matter, and
4\% atoms \cite{WMAP7}. Currently, several DM search experiments
have found possible DM signals. The indirect DM detection
experiments PAMELA \cite{PAMELA}, Fermi \cite{Fermi} and ATIC
\cite{ATIC} have observed the cosmic electron anomalies which can be
explained by annihilating or decaying DM models \cite{Guo:2010vy}.
The direct DM detection experiment CDMS II \cite{CDMSII}  observed
two possible events in the signal region. In addition, the
DAMA/LIBRA \cite{DAMA} and CoGeNT \cite{CoGeNT} results favor a
light DM candidate with a very large DM-nuclei elastic scattering
cross section.


In the standard model (SM) of particle physics, there is no
candidate for dark matter. Therefore, one has to extend the SM to
account for the dark matter. It is well known that the simplest DM
model can be constructed by adding a real singlet scalar $S$ to the
standard model. In this model, a discrete $Z_2$ symmetry ($S
\rightarrow -S$) has to be introduced to make the DM candidate $S$
stable. Although this model is very simple, it is phenomenologically
interesting. Motivated by the simplicity and predictability, a
number of authors have explored its phenomenology
\cite{McDonald,Burgess:2000yq,SingletDM,Barger:2007im,Goudelis:2009zz,Gonderinger:2009jp,DAMACoGeNT}.
These research works are very helpful for us to understand some more
complicated DM models.


In this paper, we try to give a comprehensive analysis for the real
singlet scalar DM model. Instead of assuming the fixed Higgs mass,
we scan the whole parameter space of the DM and Higgs masses. For
the resonance case, one should consider the Breit-Wigner resonance
effect. Once the coupling between DM particle and Higgs boson is
derived from the observed DM abundance, the DM elastic scattering
cross section on a nucleon and the DM annihilation cross section in
the galactic halo can be straightly calculated. The current DM
direct and indirect detection experiments can be used to constrain
the model parameter space. In addition, we also discuss the Higgs
searches at colliders when the Higgs can decay into two DM
particles. This paper is organized as follows: In Sec. II, we
outline the main features of this model and give the DM annihilation
cross sections. In Sec. III, we investigate the DM direct search,
the DM indirect search and  the collider implications. Some
discussions and conclusions are given in Sec. IV.


\section{The real singlet scalar dark matter model}

In the real singlet scalar DM model, the Lagrangian reads
\begin{eqnarray}
\mathcal{L} = \mathcal{L}_{\rm SM} + \frac{1}{2}\partial_\mu S
\partial^\mu S - \frac{m_0^2}{2} S^2 - \frac{\lambda_S}{4} S^4 - \lambda S^2 H^\dag H \; , \label{L}
\end{eqnarray}
where $H$ is the SM Higgs doublet. The linear and cubic terms of the
scalar $S$ are forbidden by the $Z_2$ symmetry $S \rightarrow -S$.
Then $S$ has a vanishing vacuum expectation value (VEV) which
ensures the DM candidate $S$ stable. $\lambda_S$ describes the DM
self-interaction strength which is independent of the DM
annihilation. The observations of galactic DM halos and the dynamics
of the Bullet cluster may constrain $\lambda_S$ when the DM mass is
the order of 1 to 100 MeV \cite{McDonald:2007ka}. Notice that the
one-loop vacuum stability can lead to a lower bound on the DM mass
for a given $\lambda_S$ \cite{Gonderinger:2009jp}. It is clear that
the DM-Higgs coupling $\lambda$ is the only one free parameter to
control the DM annihilation. After the spontaneous symmetry
breaking, one can obtain the DM mass $m_D^2 = m_0^2 + \lambda \;
v_{\rm EW}^2$ with $v_{\rm EW}= 246$ GeV. It is natural for us to
take $m_D$ in the range of a few GeV and a few hundred GeV. In this
paper, we focus on $10 \;{\rm GeV} \leq m_D \leq 200$ GeV. In
addition, the Higgs mass $m_h$ is also an important parameter for
our numerical calculation. The $95 \%$ confidence-level (CL) lower
bound on the Higgs mass is $m_h > 114.4$ GeV given by the LEP
accelerator \cite{Barate:2003sz}. The upper limit is $m_h < 186 $
GeV when we consider the precision electroweak data and the LEP
direct lower limit \cite{Alcaraz:2009jr}. Therefore, we choose
$114.4 \; {\rm GeV} <  m_h < 186 \; {\rm GeV}$. It is worthwhile to
stress that the current Tevatron experiments CDF and D0 have
excluded $162 \; {\rm GeV}<  m_h < 166$ GeV \cite{Aaltonen:2010yv}.

The real singlet scalar DM model is very simple and has only three
free parameters $m_D$, $m_h$ and $\lambda$. Based on the DM mass
$m_D$, the DM candidate $S$ may annihilate into fermion pairs, gauge
boson pairs or Higgs pairs. The annihilation cross sections
$\hat{\sigma}= 4 E_1 E_2 \sigma v$ ($E_1$ and $E_2$ are the energies
of two incoming DM particles) for different annihilation channels
have the following forms:
\begin{eqnarray}
\hat{\sigma}_{ff} &=& \sum_f \frac{\lambda^2 m_f^2}{\pi}
\frac{1}{(s-m_{h}^2)^2+m_{h}^2 \Gamma_{h}^2}
\frac{(s-4m_f^2)^{1.5}}{\sqrt{s}}, \label{FF} \\
\hat{\sigma}_{ZZ} &=&  \frac{\lambda^2 }{4 \pi}
\frac{s^2}{(s-m_{h}^2)^2+m_{h}^2 \Gamma_{h}^2} \sqrt{1-\frac{4
m_{Z}^2}{s}} \left(1- \frac{4m_{Z}^2}{s}+ \frac{12 m_{Z}^4}{s^2}
\right),\\
\hat{\sigma}_{WW} &=&  \frac{\lambda^2 }{2 \pi}
\frac{s^2}{(s-m_{h}^2)^2+m_{h}^2 \Gamma_{h}^2} \sqrt{1-\frac{4
m_{W}^2}{s}} \left(1- \frac{4m_{W}^2}{s}+ \frac{12 m_{W}^4}{s^2}
\right),\\
\hat{\sigma}_{hh} &=&  \frac{\lambda^2 }{4 \pi} \sqrt{1-\frac{4
m_{h}^2}{s}} \left[ \left(\frac{s+ 2 m_{h}^2}{s - m_{h}^2}\right)^2
- \frac{16 \lambda v_{\rm EW}^2}{s-2 m_{h}^2} \frac{s+ 2 m_{h}^2}{s
- m_{h}^2} F(\xi) + \frac{32 \lambda^2 v_{\rm EW}^4}{(s-2
m_{h}^2)^2} \left( \frac{1}{1-\xi^2} + F(\xi)\right) \right],
\label{hh}
\end{eqnarray}
where $s$ is the squared center-of-mass energy.  The quantity $F$ is
defined as $F(\xi)\equiv\mbox{arctanh}(\xi)/\xi$ with $\xi
\equiv\sqrt{(s-4m_{h}^2)(s-4m_D^2)}/(s-2m_{h}^2)$. When the DM
annihilation nears a resonance, we should know the Higgs decay width
$\Gamma_{h}$ which is given by
\begin{eqnarray}
\Gamma_{h} & = & \frac{\sum m_f^2 }{8 \pi v_{\rm EW}^2}
\frac{(m_{h}^2 - 4 m_f^2)^{1.5}}{m_{h}^2} + \frac{m_h^3}{32 \pi
v_{\rm EW}^2} \sqrt{1- \frac{4 m_Z^2}{m_h^2}} \left(1-
\frac{4m_{Z}^2}{m_h^2}+ \frac{12 m_{Z}^4}{m_h^4}\right) \\ \nonumber
& + &  \frac{m_h^3}{16 \pi v_{\rm EW}^2} \sqrt{1- \frac{4
m_W^2}{m_h^2}} \left(1- \frac{4m_{W}^2}{m_h^2}+ \frac{12
m_{W}^4}{m_h^4}\right) + \frac{\lambda^2 v_{\rm EW}^2}{8 \pi}
\frac{\sqrt{m_h^2 - 4 m_D^2}}{m_{h}^2} \;.
\end{eqnarray}
Here we have included the decay mode $h \rightarrow S S$ if $m_h > 2
m_D$.

\section{Dark matter searches}

\subsection{Dark matter relic density}

The thermal-average of the annihilation cross section times the
relative velocity $\langle \sigma v \rangle$ is a key quantity in
the determination of the DM cosmic relic abundance. We adopt the
usual single-integral formula for $\langle \sigma v \rangle$
\cite{Edsjo:1997bg}:
\begin{eqnarray}
\langle \sigma v \rangle = \frac{1}{n_{EQ}^2} \frac{m_D}{64 \pi^4 x}
\int_{4 m_D^2}^{\infty} \hat{\sigma}(s) \sqrt{s} K_1(\frac{x
\sqrt{s}}{m_D}) d s \;, \label{cross}
\end{eqnarray}
with
\begin{eqnarray}
n_{EQ} & = & \frac{g_i}{2 \pi^2} \frac{m_D^3}{x} K_2(x) \; , \label{n} \\
\hat{\sigma}(s) & = & \hat{\sigma} \; g_i^2 \; \sqrt{1-\frac{4
m_D^2}{s}} \; , \label{sigmahat}
\end{eqnarray}
where $K_1(x)$ and $K_2(x)$ are the modified Bessel functions.
$x\equiv m_D/T$ and $g_i =1$ is the internal degrees of freedom for
the scalar dark matter $S$. Using the annihilation cross section
$\hat{\sigma}$ in Eqs. (\ref{FF}-\ref{hh}), one can numerically
calculate the thermally averaged annihilation cross section $\langle
\sigma v \rangle$ by use of the above formulas.

The evolution of the DM abundance is given by the following
Boltzmann equation \cite{KOLB}:
\begin{eqnarray}
\frac{d Y}{d x} = - \frac{x \; {\bf s}(x)}{H} \langle \sigma v
\rangle (Y^2 -Y_{EQ}^2) \; , \label{bol}
\end{eqnarray}
where $Y \equiv n/{\bf s}(x)$  denotes the dark matter number
density. The entropy density ${\bf s}(x)$ and the Hubble parameter
$H$ evaluated at $x=1$ are given by
\begin{eqnarray}
{\bf s}(x) & = & \frac{2 \pi^2 g_*}{45} \frac{m_D^3}{x^3} \;, \\ H &
= & \sqrt{\frac{4 \pi^3 g_*}{45}} \frac{m_D^2}{M_{\rm PL}} \label{h}
\;,
\end{eqnarray}
where $M_{\rm PL} \simeq 1.22 \times 10^{19}$ GeV is the Planck
energy. $g_*$ is the total number of effectively relativistic
degrees of freedom. The numerical results of $g_*$ have been
presented in Ref. \cite{APP}. Here we take the QCD phase transition
temperature to be 150 MeV.

\begin{figure}[htb]
\begin{center}
\includegraphics[width=8cm,height=7cm,angle=0]{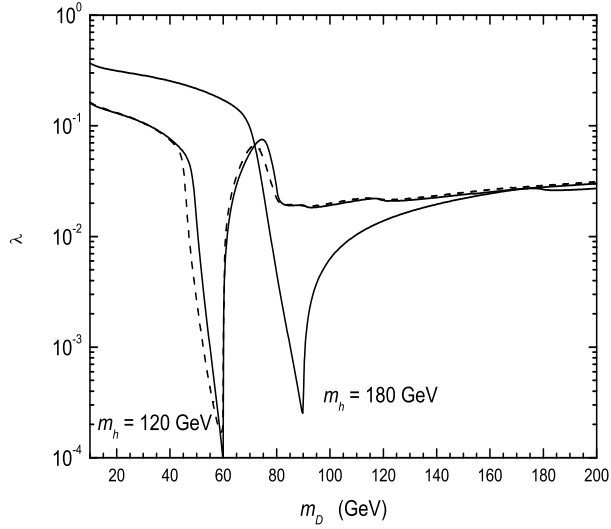}
\end{center}
\caption{ The predicted coupling $\lambda$ as a function of the DM
mass $m_D$ from the observed DM abundance for the $m_h =120$ GeV and
$m_h =180$ GeV cases. The dashed line denotes the  constant $\langle
\sigma v \rangle$ case when its value is taken at the usual
freeze-out temperature $x=20$. } \label{lambda120}
\end{figure}

Using the result $Y_0$ of the integration of Eq. (\ref{bol}), we may
obtain the DM relic density $\Omega_D h^2 = 2.74 \times 10^8 \; Y_0
\; m_D /{\rm GeV}$. In terms of the observed DM abundance $0.1088
\leq\Omega_D h^2 \leq 0.1158$ \cite{WMAP7}, one can calculate the
DM-Higgs coupling $\lambda$ for the given $m_D$ and $m_h$. As shown
in Fig. \ref{lambda120}, the observed DM abundance requires $\lambda
\sim {\cal O} (10^{-4} - 10^{-1})$. It is well known that the
annihilation cross section $\sigma$ will become larger for the same
coupling when the annihilation process nears a resonance. This
feature implies that there is a very small coupling when $0.8 \; m_h
\lesssim 2 m_D < m_h$. This region is named as the resonance region
in the following parts of this paper. It should be mentioned that
the thermally averaged annihilation cross section $\langle \sigma v
\rangle$ will significantly change as the evolution of the Universe
when the DM particle is nearly one-half the mass of a resonance
\cite{BW}. This is the Breit-Wigner resonance effect which has been
used to explain the recent PAMELA, ATIC and Fermi anomalies. Here we
have considered the Breit-Wigner resonance effect for the
determination of the coupling $\lambda$. If we neglect the
Breit-Wigner resonance effect and only consider the resonance
contribution, the predicted coupling $\lambda$ will has distinct
differences with the previous results. For illustration, we also
present the corresponding $\lambda$ in Fig. \ref{lambda120}  by use
of the constant $\langle \sigma v \rangle$ at $x =20$. In this case,
the predicted $\lambda$ may differ from the correct one by more than
a factor of 3 for the resonance region. Since the DM direct and
indirect detection cross sections are proportional to $\lambda^2$,
one can derive the larger differences.

\subsection{Dark matter direct search}

For the scalar dark matter, the DM elastic scattering cross section
on a nucleus ${\cal N}$ is spin-independent, which is given by
\cite{DM}
\begin{eqnarray}
\sigma_{\cal N} = \frac{4 M^2({\cal N})}{\pi} (Z f_p + (A-Z) f_n)^2
\;,
\end{eqnarray}
where $M({\cal N}) = m_D M_{\cal N}/(m_D + M_{\cal N})$ and $M_{\cal
N}$ is the target nucleus mass. $Z$ and $A-Z$ are the numbers of
protons and neutrons in the nucleus. $f_{p,n}$ is the coupling
between DM and  protons or neutrons, given by
\begin{eqnarray}
f_{p,n}= \sum_{q=u,d,s} f_{Tq}^{(p,n)} a_q \frac{m_{p,n}}{m_q} +
\frac{2}{27} f_{TG}^{(p,n)}  \sum_{q=c,b,t} a_q \frac{m_{p,n}}{m_q},
\label{fn}
\end{eqnarray}
where $f_{Tu}^{(p)}=0.020 \pm 0.004$, $f_{Td}^{(p)}=0.026 \pm
0.005$, $f_{Ts}^{(p)}=0.118 \pm 0.062$, $f_{Tu}^{(n)}=0.014 \pm
0.003$, $f_{Td}^{(n)}=0.036 \pm 0.008$ and $f_{Ts}^{(n)}=0.118 \pm
0.062$ \cite{Ellis:2000ds}. The coupling $f_{TG}^{(p,n)}$ between DM
and gluons from heavy quark loops is obtained from $f_{TG}^{(p,n)} =
1 - \sum_{q=u,d,s} f_{Tq}^{(p,n)}$, which leads to $f_{TG}^{(p)}
\approx 0.84 $ and $f_{TG}^{(n)} \approx 0.83$. The results of
DM-nucleus elastic scattering experiments are presented in the form
of a normalized DM-nucleon scattering cross section
$\sigma_{n}^{SI}$ in the spin-independent case, which is
straightforward
\begin{eqnarray}
\sigma_{n}^{SI} =\left (\frac{m_D \; m_n}{m_D+ m_n } \right)^2
\frac{1}{A^2 M^2({\cal N})} \sigma_{\cal N} \approx \frac{4}{\pi}
\left (\frac{m_D \; m_n}{m_D+ m_n } \right)^2 f_n^2 \; ,
\end{eqnarray}
where we have used $f_p \approx f_n$.

\begin{figure}[htb]
\begin{center}
\includegraphics[width=8cm,height=7cm,angle=0]{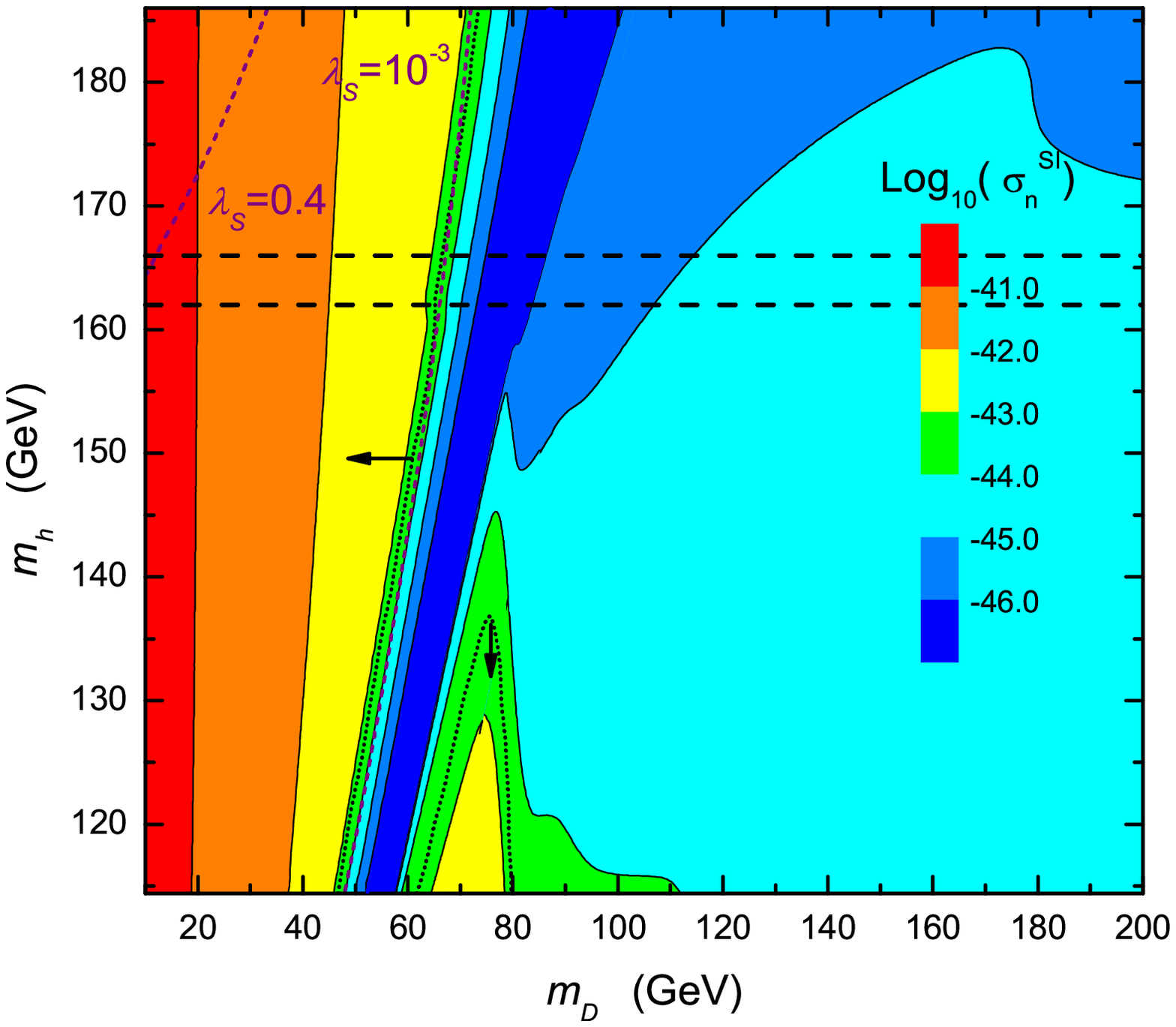}
\includegraphics[width=8cm,height=7cm,angle=0]{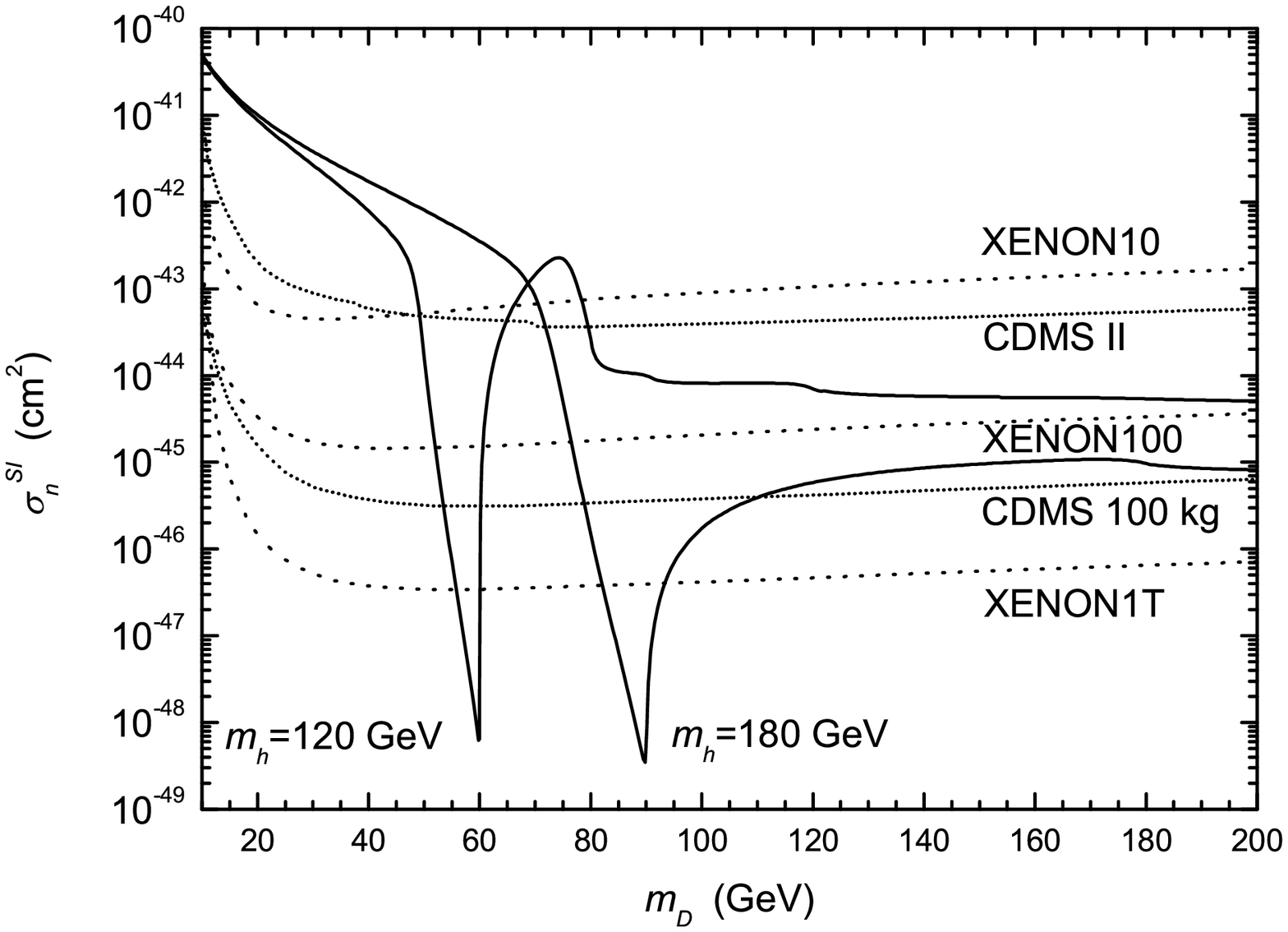}
\end{center}
\caption{ The predicted DM-nucleon scattering cross section
$\sigma_{n}^{SI}$ for the whole parameter space. The short dotted
lines with arrowhead indicate the excluded regions from the DM
direct search experiments CDMS II and XENON10. The region among two
dashed lines ($162 \; {\rm GeV}<  m_h < 166$ GeV) can be excluded by
the Tevatron experiments CDF and D0. Two purple short dashed lines
describe the minimum $m_D$ allowed by both the DM observed abundance
and the vacuum stability/pertubativity for $\lambda_S = 10^{-3}$ and
$\lambda_S = 0.4$. The right panel corresponds to two fixed Higgs
mass cases with current and future experimental upper bounds. }
\label{Direct}
\end{figure}

In the real singlet scalar DM model, the DM candidate $S$ interacts
with nucleus ${\cal N}$ through  $t$-channel Higgs exchange. In this
case, the DM-quark coupling $a_q$ in Eq. (\ref{fn}) is given by
\begin{eqnarray}
a_q = \frac{\lambda \; m_q}{m_D \; m_h^2}  \,. \label{aq}
\end{eqnarray}
Using the predicted $\lambda$ from the observed DM abundance,  one
can calculate the spin-independent DM-nucleon elastic scattering
cross section $\sigma_{n}^{SI}$ for the given $m_D$ and $m_h$. We
perform a numerical scan over the whole parameter space of $m_D$ and
$m_h$. The numerical results are shown in Fig. \ref{Direct}. For
illustration, we have also plotted $\sigma_{n}^{SI}$ as a function
of $m_D$ for the $m_h = 120$ GeV and $m_h = 180$ GeV cases. In view
of the current DM direct detection upper bounds from CDMS II
\cite{CDMSII} and XENON10 \cite{XENON10}, we find that two regions
indicated by short dotted lines can be excluded. The future
experiments XENON100 \cite{XENON100P}, CDMS 100 kg \cite{CDMS100}
and XENON1T \cite{XENON1T} can cover most parts of the allowed
parameter space. For the resonance region, the predicted
$\sigma_{n}^{SI}$ is far below the current experimental upper
bounds. In Fig. \ref{Direct}, we also plot the minimum $m_D$ allowed
by both the DM observed abundance and the vacuum
stability/pertubativity for the DM self-coupling $\lambda_S =
10^{-3}$ and $\lambda_S = 0.4$ when the cutoff scale is taken to be
1 TeV \cite{Gonderinger:2009jp}. Increasing the cutoff scale, one
can derive much stronger bounds \cite{Gonderinger:2009jp}. In
addition, the region among two dashed lines ($162 \; {\rm GeV}< m_h
< 166$ GeV) can be excluded by the Tevatron experiments CDF and D0.

\subsection{Dark matter indirect search}

\begin{figure}[htb]
\begin{center}
\includegraphics[width=8cm,height=7cm,angle=0]{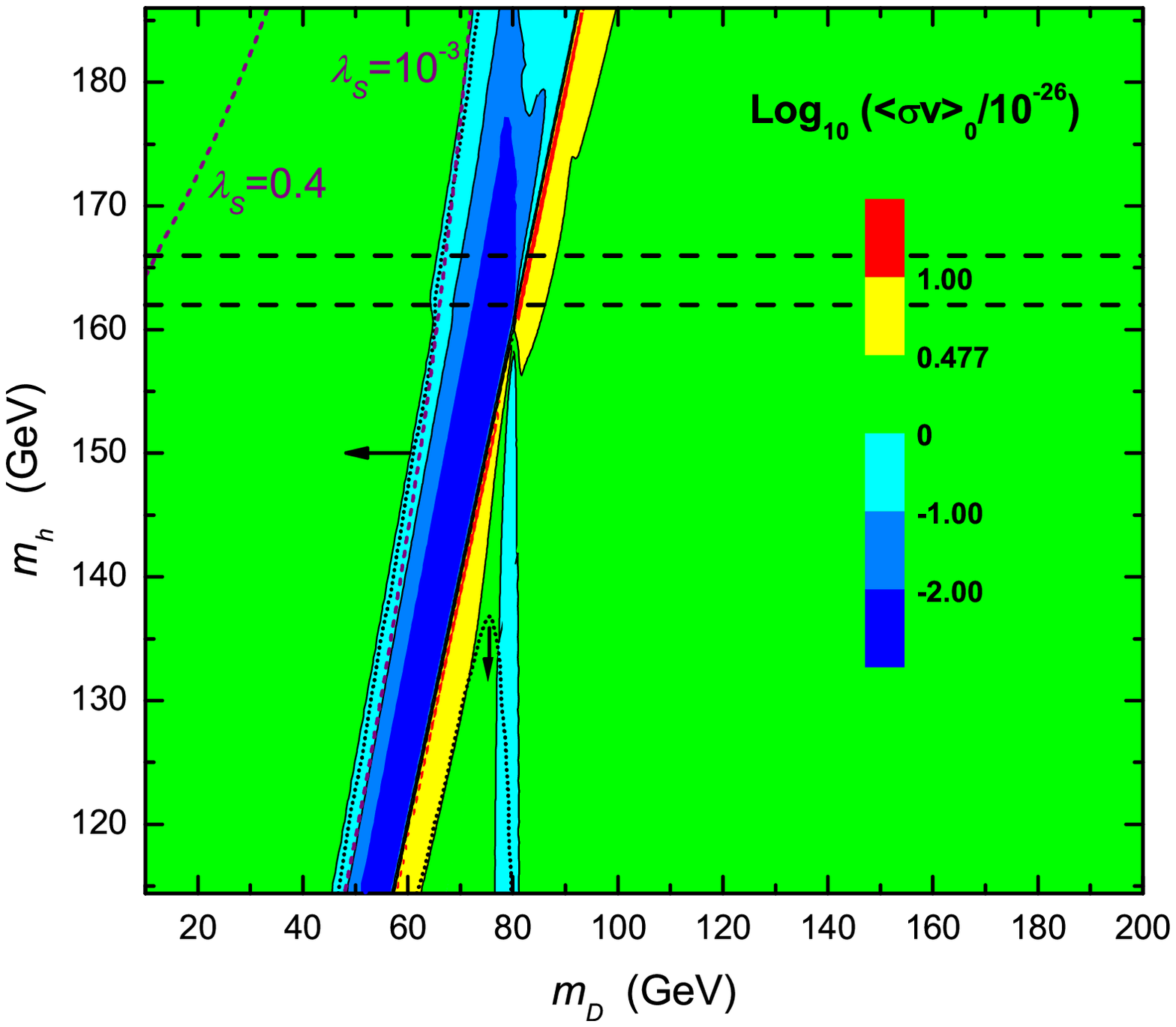}
\includegraphics[width=8cm,height=7cm,angle=0]{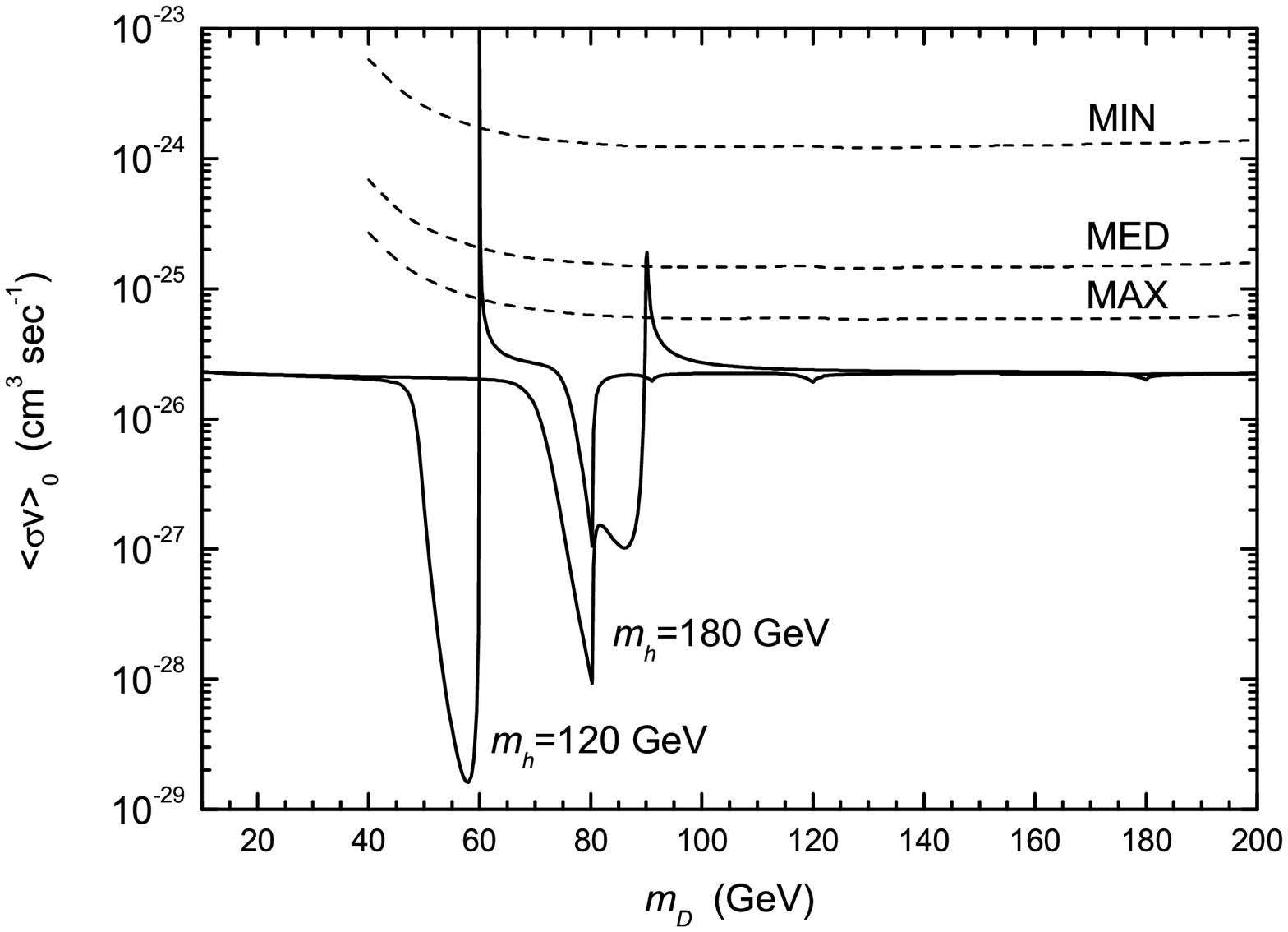}
\end{center}
\caption{  The predicted thermally averaged DM annihilation cross
section $\langle \sigma v \rangle_0$. The right panel corresponds to
two fixed Higgs mass cases. } \label{Indirect}
\end{figure}

The DM indirect search experiments are designed to detect the DM
annihilation productions which include neutrinos, gamma rays,
electrons, positrons, protons and antiprotons. Since these methods
are complementary to the direct detection, it will be very important
for us to investigate the DM indirect  detection.  In order to
derive the correct relic density, we have calculated the thermally
averaged annihilation cross section $\langle \sigma v \rangle$ which
depends on the temperature of the Universe. On the other hand,
$\langle \sigma v \rangle$ also determines the DM annihilation rate
in the galactic halo. The only difference among the above two cases
is the temperature $T$. For the relic density, $\langle \sigma v
\rangle$ is usually evaluated at the freeze-out temperature $x
\approx 20$ (the averaged velocity $v \approx \sqrt{3/x}$). The dark
matter annihilation in the galactic halo occurs at $x \approx 3
\times 10^6$ ($v \approx 10^{-3}$). Therefore we should calculate
the thermally averaged annihilation cross section at $x \approx 3
\times 10^6$, namely $\langle \sigma v \rangle_0$. The numerical
results have been shown in Fig. \ref{Indirect}. One may find $1
\times 10^{-26} \;{\rm cm}^3 \; {\rm sec}^{-1} \leq\langle \sigma v
\rangle_0 \leq 3 \times 10^{-26} \;{\rm cm}^3 \; {\rm sec}^{-1}$ for
most parts of the parameter space, which is consistent with the
usual $s$-wave annihilation cross section $\langle \sigma v \rangle
\approx 3 \times 10^{-26} \;{\rm cm}^3  \; {\rm sec}^{-1}$ at the
freeze-out temperature $x \approx 20$. However, the Breit-Wigner
resonance effect can enhance or suppress $\langle \sigma v \rangle$
for the resonance case \cite{BW}.  As shown in Fig. \ref{Indirect},
$\langle \sigma v \rangle$ in the red and yellow regions is enhanced
by the Breit-Wigner enhancement mechanism. Contrarily, $\langle
\sigma v \rangle$ is suppressed ($\langle \sigma v \rangle \ll 1
\times 10^{-26} \;{\rm cm}^3  \; {\rm sec}^{-1}$) when double DM
mass $2 m_D$ is slightly less than the Higgs mass $m_h$. In this
case, it is very difficult for us to detect the signals of the DM
annihilation. The vertical cyan region around 80 GeV has smaller
$\langle \sigma v \rangle_0$ which originates from the threshold
effect \cite{Threshold}. If $m_D$ is slightly less than the $W$
boson mass, the channel $S S \rightarrow W^+ W^-$ is open at high
temperature. It dominates the total thermally averaged annihilation
cross section and determines the DM relic density. However, this
channel is forbidden in the galactic halo (low relative velocity).
At this moment, the channel $S S \rightarrow b \bar{b}$ has the
dominant contribution. This feature can be well understood from Fig.
\ref{Indirect} (right panel).

In our model, the DM annihilation can generate primary antiprotons
which can be detected by the DM indirect search experiments.
Recently, the PAMELA collaboration reports that the observed
antiproton data is consistent with the usual estimation value of the
secondary antiproton \cite{PAMELA}. Therefore one can use the PAMELA
antiproton measurements to constrain $\langle \sigma v \rangle_0$.
In Fig. \ref{Indirect}, we have also plotted the maximum allowed
$\langle \sigma v \rangle_0$ for the MIN, MED and MAX antiproton
propagation models \cite{Goudelis:2009zz}. Notice that a fixed Higgs
mass $m_h =120$ GeV has been assumed in Ref. \cite{Goudelis:2009zz}.
The above upper bounds are still valid for our analysis when $m_D <
114.4$ GeV. Then we can find that a very narrow region can be
excluded by the PAMELA antiproton data in our model. This feature is
not shown in Ref. \cite{Goudelis:2009zz}. In fact, the width of this
excluded region is about $0.4$ GeV if we require $\langle \sigma v
\rangle_0 \lesssim 10^{-25} \;{\rm cm}^3 \; {\rm sec}^{-1}$. For the
MED and MAX propagation cases, the sensitivity of the future
experiment AMS-02 \cite{AMS02} may reach $\langle \sigma v \rangle_0
\sim {\cal O} (10^{-27} - 10^{-26}) \;{\rm cm}^3 \; {\rm sec}^{-1}$
\cite{Goudelis:2009zz},  which can cover most parts of the whole
parameter space as shown in Fig. \ref{Indirect}.

\subsection{Implications on the Higgs search}

\begin{figure}[htb]
\begin{center}
\includegraphics[width=8cm,height=7cm,angle=0]{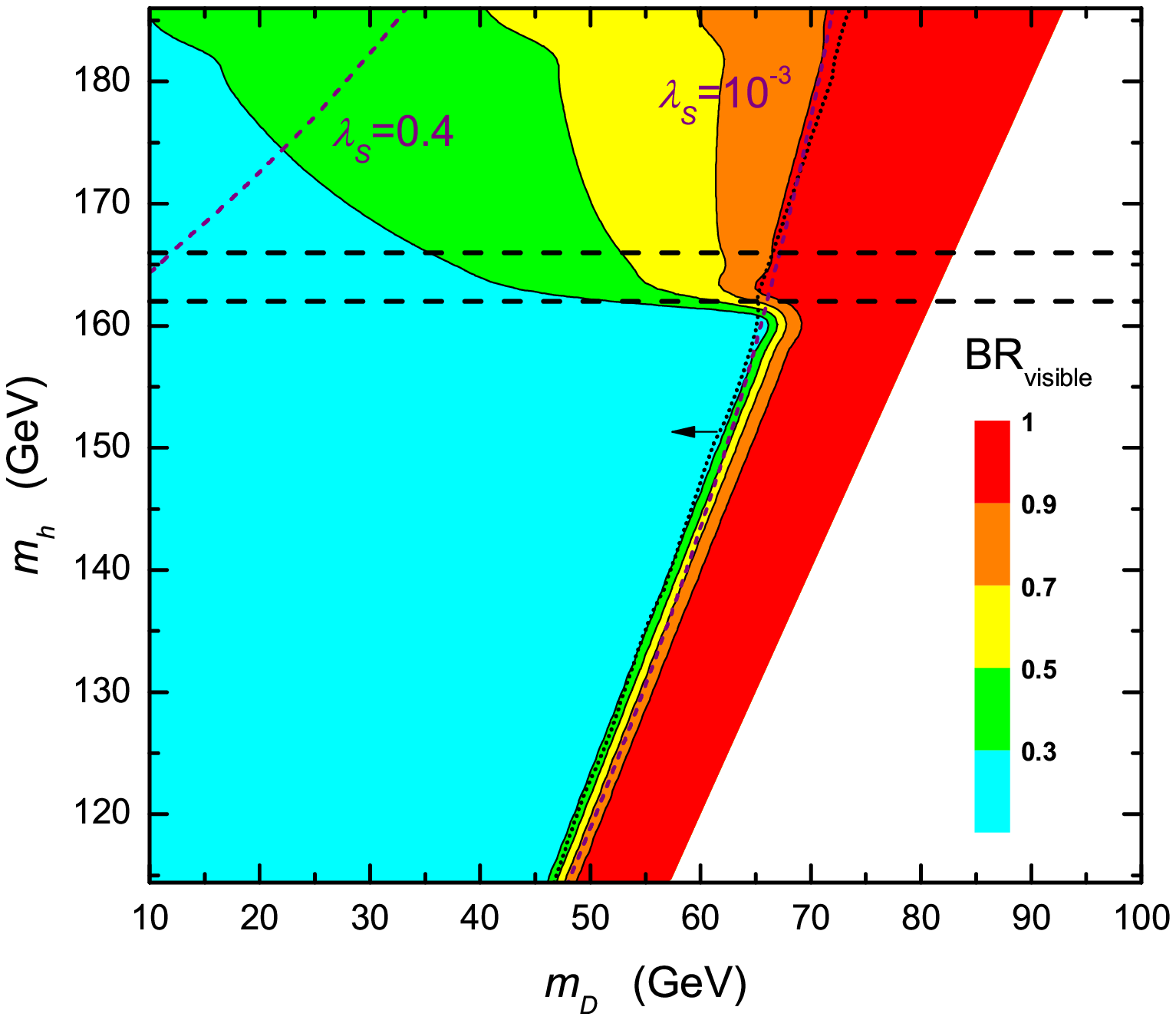}
\includegraphics[width=8cm,height=7cm,angle=0]{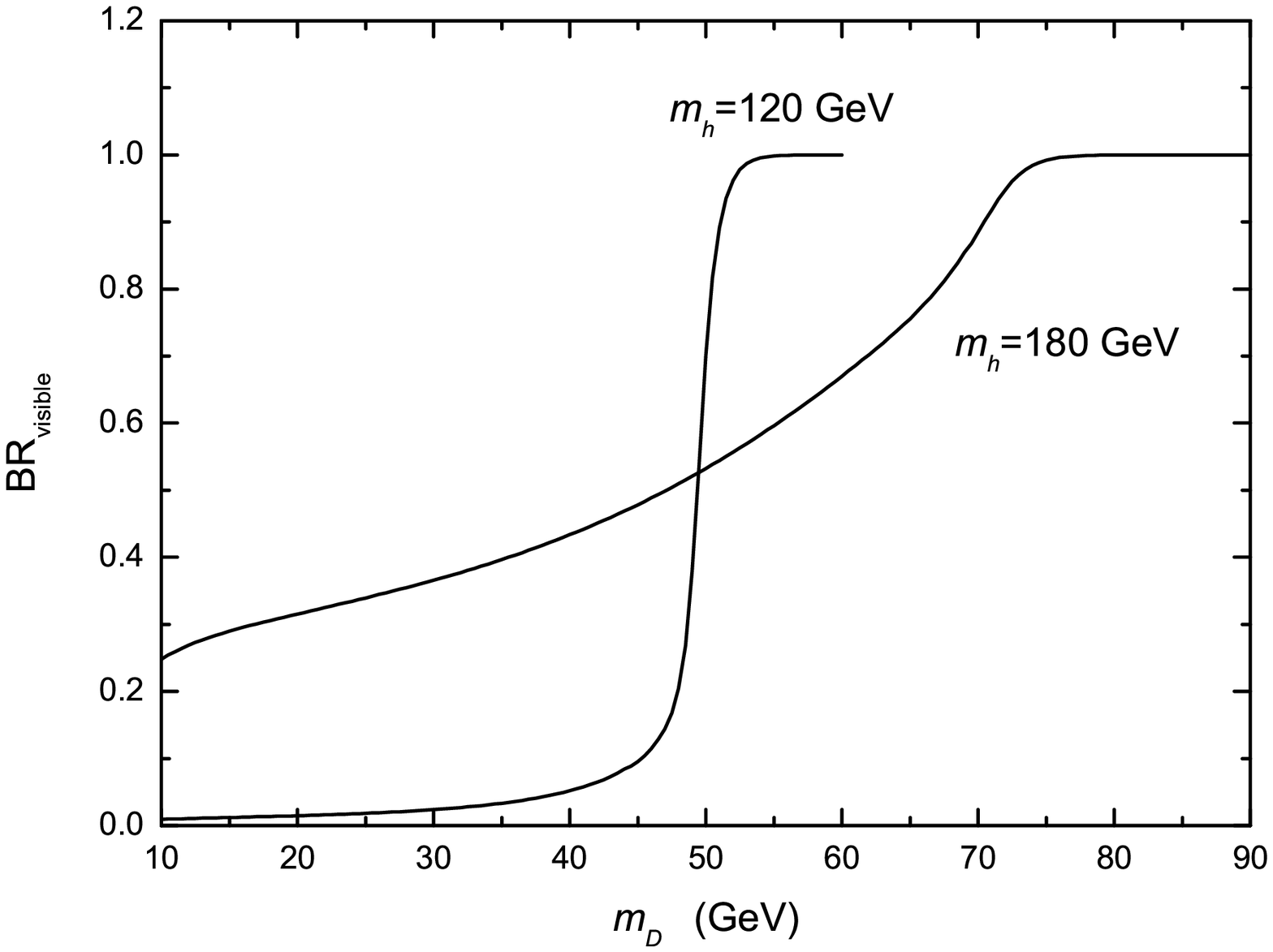}
\end{center}
\caption{ The predicted branching ratio of the Higgs visible decay
${\rm BR_{visible}}$. The right panel corresponds to two fixed Higgs
mass cases. } \label{BR}
\end{figure}

Since the DM candidate $S$ has substantial coupling to nucleons via
Higgs boson exchange, they can be produced in high energy collider
experiments such as the Tevatron and Large Hadron Collider (LHC). If
$m_h > 2 m_D$, the decay channel $h \rightarrow 2S$ is kinematically
allowed. In this case, two DM particle production may be associated
with the Higgs production. The produced DM particles are invisible
and have the missing energy signals. This will affect the usual SM
Higgs searches at the Tevatron and LHC. To describe this effect, we
calculate the branching ratio of the Higgs visible decay
\cite{Burgess:2000yq}
\begin{eqnarray}
{\rm BR_{visible}} = \frac{\Gamma_{h\rightarrow{\rm
SM}}}{\Gamma_{h\rightarrow 2S}+ \Gamma_{h\rightarrow{\rm SM}}} \; .
\end{eqnarray}
The numerical results have been shown in Fig. \ref{BR}. Considering
the constraints from the DM direct search experiments, we find ${\rm
BR_{visible}} \gtrsim 0.3$.  In fact, we have ${\rm BR_{visible}} >
0.9$ for most parts of the allowed parameter space. If the future
CDMS 100 kg does not observe the DM signals, the region less than
0.9 can be excluded. Notice that the invisible Higgs decays in the
allowed region $0.3 \lesssim {\rm BR_{visible}} \lesssim 0.75$ can
be identified at the ATLAS with an integrated luminosity of 30
fb$^{-1}$ \cite{Warsinsky:2008zza}. The decay mode $h \rightarrow S
S$ can lower the statistical significance of the traditional Higgs
search at the CMS \cite{Barger:2007im}. A combined analysis of the
traditional visible search and the invisible search at the LHC can
enhance the discovery reach and the possibility of distinguishing
this DM model from the SM \cite{Barger:2007im}.

\section{Discussion and Conclusion}

We have detailedly discussed the $10 \;{\rm GeV} \leq m_D \leq 200$
GeV case. As shown in Fig. \ref{Direct}, one can obtain the very
large DM-nucleon elastic scattering cross section
$\sigma_{n}^{SI}\sim {\cal O} (10^{-41} {\rm cm}^2)$ for $m_D \sim
10$ GeV. In this case, the light DM particle $S$ can explain the
DAMA/LIBRA and CoGeNT experiments \cite{DAMACoGeNT}. It should be
mentioned that this explanation is consistent with the XENON10 and
CDMS null results \cite{Petriello:2008jj}. Currently, the XENON100
preliminary results challenge the DM interpretation of the
DAMA/LIBRA and CoGeNT signals \cite{Aprile:2010um}. However, there
are bifurcations on the choice of the ratio between electron
equivalent energy and nuclear recoil energy \cite{Collar:2010gg}. If
$m_D > 200$ GeV, one will not meet the resonance and the new
annihilation channels. Then $\langle \sigma v \rangle_0 \approx 2.3
\times 10^{-26} \;{\rm cm}^3 \; {\rm sec}^{-1}$ and $10^{-45} \;
{\rm cm}^2 \lesssim \sigma_{n}^{SI} \lesssim 10^{-44} \; {\rm cm}^2$
can be derived.

In conclusion, we have made an undated comprehensive analysis for
the whole parameter space of the real singlet scalar DM model. To
satisfy the observed DM abundance, we numerically solve the
Boltzmann equation and predict the DM-Higgs coupling $\lambda$ which
determines the DM direct and indirect detection rates. We
demonstrate that the Breit-Wigner resonance effect can significantly
change the predicted coupling $\lambda$ by more than a factor of 3
for the resonance region. The spin-independent DM-nucleon elastic
scattering cross section $\sigma_{n}^{SI}$ has been presented for
the whole parameter space of $m_D$ and $m_h$. One may find that the
current experimental upper bounds from CDMS II and XENON10 can
exclude two regions. For the DM indirect detection, we calculate the
thermally averaged annihilation cross section $\langle \sigma v
\rangle_0$ which can be enhanced or suppressed by the Breit-Wigner
resonance effect. We find that a very narrow region can be excluded
by the PAMELA antiproton measurements. One should notice that the
predicted $\sigma_{n}^{SI}$ and $\langle \sigma v \rangle_0$ are
very small for the resonance region. In this case, it is very
difficult for us to detect the signals of the DM annihilation.
However, we still have possibility to test the resonance region as
detector masses of DM direct search experiments become larger. When
the decay channel $h \rightarrow S S$ is kinematically allowed, we
find that the allowed branching ratio of the Higgs visible decay
satisfy ${\rm BR_{visible}} \gtrsim 0.3$. If the future CDMS 100 kg
does not observe the DM signals, the region less than 0.9 can be
excluded.

\acknowledgments

This work is supported in part by the National Basic Research
Program of China (973 Program) under Grants No. 2010CB833000; the
National Nature Science Foundation of China (NSFC) under Grants No.
10821504 and No. 10905084; and the Project of Knowledge Innovation
Program (PKIP) of the Chinese Academy of Science.

\end{document}